\documentclass[letter]{article}
  \usepackage[totalwidth=420pt,totalheight=625pt]{geometry}
  \date{}
\usepackage{mycitations}  
\usepackage{caption}

\usepackage{hyperref}
\usepackage{amsmath}
\usepackage{amsthm}
\usepackage{cleveref}
\usepackage{caption}

\usepackage{algorithm}
\usepackage{algorithmic}
\usepackage{amssymb,amsmath,amsthm}
\usepackage{enumitem}
\usepackage{newfloat}
\usepackage{cleveref}
\usepackage{listings}
\usepackage[most]{tcolorbox}
\tcbuselibrary{listings}

\floatstyle{ruled}
\newfloat{listing}{tb}{lst}{}
\floatname{listing}{Listing}

\usepackage{soul}   
\usepackage{fvextra}
\usepackage{booktabs}

\title{Generating Streamlining Constraints with\\ Large Language
    Models}

   \author{%
    Florentina Voboril, Vaidyanathan Peruvemba Ramaswamy, Stefan Szeider\\[4pt]
\small Algorithms and Complexity Group\\[-3pt]
\small TU Wien, Vienna, Austria\\[-3pt]
\small \texttt{\{fvoboril,vaidyanathan,sz\}@ac.tuwien.ac.at}}

\usepackage[textsize=scriptsize]{todonotes} 
\usepackage{xspace}
\newcommand{\vai}[1]{\todo{{Vaidyanathan: #1}}}
\newcommand{\stefan}[1]{\todo{{Stefan: #1}}}
\newcommand{\flor}[1]{\todo{{Florentina: #1}}}

\newcommand{\red}[1]{\textcolor{red!80!black}{#1}}
\renewcommand{\vai}[1]{}\renewcommand{\stefan}[1]{}\renewcommand{\flor}[1]{}\renewcommand{\red}[1]{}

\newcommand{\RTSLLM}{StreamLLM\xspace}
\newcommand{\minizinc}{MiniZinc\xspace}
\newcommand{\singleshot}{static\xspace}
\newcommand{\multishot}{adaptive\xspace}

\newcommand{\Multishot}{Adaptive\xspace}

\newcommand{\SSS}{\mathcal S}
\newcommand{\MMM}{\mathcal M}
\newcommand{\III}{\mathcal I}
\newcommand{\TTT}{\mathcal T}
\newcommand{\PPP}{\mathcal P}
\newcommand{\LLL}{\mathcal L}

\newcommand{\CSfull}{Car Sequencing\xspace}
\newcommand{\CS}{CS\xspace}  
\newcommand{\HCfull}{Hypergraph Coloring\xspace}
\newcommand{\HC}{HC\xspace}  
\newcommand{\BIBDfull}{Balanced Incomplete Block Design\xspace}
\newcommand{\BIBD}{BIBD\xspace}  
\newcommand{\BHfull}{Black Hole\xspace}
\newcommand{\BH}{BH\xspace}  
\newcommand{\CCfull}{Carpet Cutting\xspace}
\newcommand{\CC}{CC\xspace}  
\newcommand{\SGfull}{Social Golfers\xspace}
\newcommand{\SG}{SG\xspace}  
\newcommand{\VLfull}{Vessel Loading\xspace}
\newcommand{\VL}{VL\xspace}  

\newcommand{\myitembegin}[1]{\noindent\textbf{#1}}

\graphicspath{{NewFigures/}}

\begin{document}
\maketitle

\begin{center}
    This is the accepted version of the article published in 
    the Journal of Artificial Intelligence Research (JAIR), 2025. The final version is available on the JAIR website.
\end{center}

\thispagestyle{empty}

\begin{abstract}
Streamlining constraints (or streamliners, for short) narrow the search space,
enhancing the speed and feasibility of solving complex constraint
satisfaction problems. Traditionally, streamliners were crafted manually or
generated through systematically combined atomic constraints with high-effort
offline testing. Our approach utilizes the generative capabilities of Large Language
Models (LLMs) to propose effective streamliners for problems specified in the
MiniZinc constraint programming language and integrates feedback to the LLM
with quick empirical tests for validation. Evaluated across seven diverse
constraint satisfaction problems, our method achieves substantial runtime
reductions. We compare the results to obfuscated and disguised variants of
the problem to see whether the results depend on LLM memorization. We also
analyze whether longer offline runs improve the quality of streamliners and
whether the LLM can propose good combinations of streamliners. 
\end{abstract}

\section{Introduction}

\emph{Streamliners} %
are certain constraints added to a
constraint model to reduce the search space, thereby improving the
feasibility and speed of finding solutions to complex constraint
  satisfaction problems. By incorporating domain-specific knowledge,
streamliners can guide the constraint solver, allowing it to bypass
less promising areas of the search space. Gomes and
  Sellmann~\shortcite{GomesSellmann04} introduced streamliners to speed up
  the constrained-based search for hard combinatorial design
  problems. Today, streamliners are a standard tool for speeding up
constrained-based search. Streamliners are closely related to
implied/redundant constraints, symmetry-breaking constraints, and
dominance-breaking constraints; however, adding a streamliner may even
cause the constraint model to become inconsistent.

Originally, streamliners were hand-crafted by researchers who used
their theoretical insight to analyze the constrained model. However,
progress has also been made on the automated generation of
streamliners
\cite{SpracklenDAM23}
by systematically trying the effect of some atomic constraints, such as
imposing specific constraints on integer and function domains, like enforcing
odd or even values, monotonicity, and properties like commutativity, as well as
facilitating specific attributes in binary relations. These atomic restrictions
are tested on thousands of problem instances, and those that show a good
streamlining effect are systematically combined.

This paper proposes a different approach to automated streamliner
generation that utilizes the generative capabilities (``creativity'') of Large Language Models
(LLMs) to generate streamliners, providing a neuro-symbolic prototype
implementation \RTSLLM, and rigorously tests it on seven different
constraint satisfaction problems using 
at least 310 CPU days\footnote{This is a
  conservative estimate; we spent about 24 CPU days for instance
  generation, 75 CPU days for grading the instances, 162 CPU days
  for evaluating the performance of the generated streamliners,
  20 CPU days for the offline approach, and 28 days for the
  combinations approach.}.  Our
approach leverages the capabilities of LLMs to infer potentially
effective streamliners, similar to an experienced researcher's
intuitive grasp of a problem. By integrating LLMs into the streamliner
generation process, we can potentially uncover unique and subtle
patterns in the problem formulation and utilize them to speed up  
the solving process.  To some extent, \RTSLLM is closer related to a
hand-crafted streamliner design by human experts rather than an automated
bottom-up streamliner generation, the dominant strategy of the previous
research~\cite{SpracklenDAM23}.

Our system \RTSLLM  for streamliner generation combines queries to LLMs with
quick empirical validation on small test instances solvable within seconds. The
system generates effective constraints within minutes through engineered prompts
and adaptive feedback based on streamliner performance. This realtime approach
allows streamliner benefits to be realized even for hard problem instances that
would typically require hours to solve---any significant speedup can outweigh
the brief time spent on streamliner generation.

We evaluate \RTSLLM extensively across seven diverse constraint problems,
including standard benchmarks and a novel hypergraph coloring problem, using
hundreds of instances specified in MiniZinc. The problems are formulated in the
MiniZinc constraint programming language~\cite{NethercoteSBBDT07}. We test
\RTSLLM with two state-of-the-art LLMs (GPT-4o and Claude~3.5 Sonnet), different
prompting variants, and different strategies for combining \multishot prompting
with test runs on easy instances.   Our results demonstrate remarkable
effectiveness, with some streamliners reducing solving time by up to 99\%.

To investigate whether these improvements stem from memorization or genuine
problem understanding, we test \RTSLLM  on disguised and obfuscated variants of
the problems. The system's continued strong performance, particularly on
problems with no known streamliners in the literature, suggests real analytical
capabilities rather than mere pattern matching.

Beyond the realtime setting, we explore offline approaches with extended
training periods and larger instance sets, achieving even better results in some
cases. We also investigate \RTSLLM's ability to generate effective combinations
of streamliners, finding that this often outperforms single constraints.
Analysis of the generated streamliners reveals an interesting mix---some mirror
expert-designed constraints while others take novel yet highly effective
approaches. 

Our experimental results suggest that LLM-generated streamliners can
meaningfully reduce runtime across diverse constraint satisfaction problems with
minimal problem-specific tuning, pointing to promising applications of AI
assistance in constraint programming for computationally challenging problems. 

\medskip\noindent Part of this work appeared in preliminary and shortened form at the 1st International Workshop on Neuro-Symbolic Software Engineering (NSE)~\cite{voboril_streamllm_2025}. 

\medskip\noindent\emph{The source code and benchmark instances can be found on Zenodo~\cite{Zenodo25}.}

\section{Preliminaries}
\label{sec:pre}

\subsection{Constraint Programming}


\emph{Constraint Programming}~(CP) is a methodology for solving combinatorial
problems specified by means of declarative constraints. Please refer to Rossi,
Van Beek, and Walsh~\shortcite{RossivanBeekWalsh06} for a comprehensive
discourse. Examples include scheduling, routing, planning, etc. These problems
can be of two types--\emph{decision problems} requiring a yes/no answer, while
\emph{optimization problems} requiring you to find a solution that minimizes or
maximizes a given objective function. In this paper, we focus only on decision
problems.

\emph{\minizinc}~\cite{NethercoteSBBDT07} is a popular tool for solving CP
problems. Problem specifications are written in the \minizinc modeling language
and are called \emph{models}. \minizinc compiles the model to a lower-level
language and then solves it using one of many underlying solvers like Chuffed,
Gecode, OR-Tools etc. Similarly, \emph{Conjure}~\cite{AkgunEtal11,AkgunEtal22}
is also a tool for higher-level constraint programming, which uses
\emph{Essence} as its modeling language.
In this paper, we use \minizinc as the modeling language of choice, while
Conjure is used in the instance generation pipeline AutoIG. More specifically,
the structure and format of the instances are specified in Essence and given to
Conjure, which then generates instances following these constraints.

\subsection{Streamliners}
\label{sec:streamliners}

\emph{Streamlining constraints} or \emph{streamliners}, introduced by
Gomes and Sellmann~\shortcite{GomesSellmann04},
are constraints that are added to the constraint programming models with the
goal of narrowing the focus to a small but
highly promising segment of the search space. Thus, they have the
potential to significantly reduce the search effort for hard
combinatorial problems. For example,
Gomes and Sellmann~\shortcite{GomesSellmann04} used
constraints for ``Latin squareness'' as a streamliner for finding
Diagonally Ordered Magic Squares. Streamliners have been successfully
applied in a diverse range of settings, such as combinatorial design~\cite{DiazBG17,LeBrasGS12,SmithGF05}, Ramsey-type problems and discrepancy~\cite{LiuCH21,LeBrasGS14}, SAT/CSP solver design~\cite{GomesSS06,GroverAE18,AnsoteguiMOST22,HeuleKS19}, and automatically generated streamliners~\cite{SpracklenDAM,SpracklenDAM23}.

We note that streamlining constraints are not required to be
sound. This means that adding the streamlining constraint may make a
satisfiable instance of a constraint model unsatisfiable. As a
consequence, an unsatisfiable instance in the streamlined model does
not imply that the instance is unsatisfiable in the original model.

Other constraints that are similar to streamliners are \emph{implied} (or
\emph{redundant}) constraints, \emph{symmetry-breaking} constraints,
and \emph{dominance-breaking} constraints. Implied constraints do not
change the set of feasible solutions~\cite{FrischJM04,FrischMW01,CharnleyCM06,%
ColtonM01,FrischMW03}.
Symmetry-breaking
constraints eliminate certain solutions within each equivalence class
while ensuring that at least one solution from each class
remains~\cite{FlenerFHKMPW01,FrischHKMW06,FrischJMM07,FrischHKMW09,ItzhakovC22,%
FichteHS20a}.
Dominance-breaking constraints are applicable in the context of
optimization problems (possibly formulated as a decision problem with
the objective value explicitly stated in the model), as they disallow
solutions that are known to be suboptimal. It might disallow optimal
solutions as well, as long as at least one optimal solution
remains~\cite{PrestwickB04,ChuS15,LeeZ22,GomesSellmann04}.
In contrast to streamliners, such constraints are guaranteed to be sound.

Finding a useful streamliner manually by a human expert is a
time-consuming process.  Therefore, it is appealing to generate
streamliners automatically, a direction explored successfully in
previous
work~\cite{WetterAM15,SpracklenAM18,SpracklenDAM19,SpracklenDAM23}.
A common point between all these previous approaches is that they treat the
streamliner generation as a high-effort, offline task, taking up to 4 CPU days
for a single problem. The streamliner is built up systematically from
elementary constraints to reduce the domain of a decision
variable and tested on a large set of automatically generated
instances.

Similar to other recent studies \cite{SpracklenDAM23}, we use the term
streamliner in a wider sense to also accommodate redundant, symmetry-breaking
constraints. This is particularly useful in our context since, for
automatically constructed constraints, it is difficult to determine which type the new constraint is. At the same time, we acknowledge that the practical impact of different constraint types can vary substantially. Therefore, rather than focusing on a strict theoretical classification, we adopt a pragmatic viewpoint and evaluate each generated constraint empirically.


\subsection{Large Language Models (LLMs)}

Large Language Models (LLMs) are advanced AI systems based on the transformer
models \cite{VaswaniSPUJGKP17} and trained on vast data sets to produce
human-like text and source code in response to instruction
prompts~\cite{MinaeeEtal24}. These models, trained on vast amounts of text data,
can produce human-like text across various domains and
styles~\cite{BrownMRSKDNSSAA20}. Recent advancements have expanded their
capabilities beyond traditional language tasks, including generating, analyzing,
and debugging code across multiple programming
languages~\cite{ChenEtal21,Xu0NH22,WuBarrettNarodytska23,PeiBSSY23}. In addition
to code-related tasks, LLMs have shown promise in mathematical reasoning and
problem-solving \cite{LampleLLRHLEM22,RomeraParedesBNBKDREWFKF24}; they can
process and generate mathematical expressions, solve equations, and even assist
with proofs. However, it is essential to note that while LLMs can produce
seemingly correct mathematical output, their responses should be carefully
verified. The models' performance on these tasks varies, and they may sometimes
generate plausible-looking but incorrect solutions~\cite{PoluHZBBS23}. Despite
these limitations, the potential applications of LLMs in fields such as computer
science, mathematics, and engineering are substantial and continue to expand as
the technology evolves.



\section{LLM-based Streamliner Generation and Validation}
\label{sec:real_time}

In this section, we discuss the different approaches we devise to generate and
test streamliners. The key difference between our approach and the past work by
Spracklen et al.~\shortcite{SpracklenDAM23} is that instead of deriving
streamliners by composing and combining atomic constraints, we synthesize
entire streamliners in one go. We try several variants of our approach, one
which is entirely realtime and requires no precomputation (unlike Spracklen et
al.~\shortcite{SpracklenDAM23}), another which can directly synthesize 
combinations of streamliners, and one which is offline. We evaluate our approach on
benchmark problems from the literature along with modified versions of those
problems to give insights into the strengths and weaknesses of our approach.

\subsection{Realtime Approach}

The realtime approach utilizes the generative capabilities of LLMs (usually taking only a few seconds) to propose and validate streamliners in real time. Such on-the-fly
generation is not possible with high-effort streamliner generation
that builds a streamliner systematically from elementary steps and
takes several CPU days~\cite{SpracklenDAM23}.

\newcommand{\ntest}{n_{\text{test}}}
\newcommand{\ntrain}{n_{\text{train}}}
\newcommand{\ttrain}{t_{\text{train}}}

Our system submits several queries (prompts) to the LLM that result in
candidate streamliners and tests the candidate streamliners on a small set of
$\ntrain$ easy \emph{training instances} that can be solved in less than
$\ttrain$ seconds. Rigorous testing on several problems indicates that a
relatively small number of test
instances suffices (about 15) and even small and easy test instances
(solvable in under 10 seconds with the unstreamlined model) provide a
good indication of how well streamliners will work on large and hard
instances; we lay out these experiments in more detail in
Sections~\ref{exp_training_time} and~\ref{exp_training_number}.
We test various strategies for this type of streamliner generation, which we
divide into \singleshot and \multishot categories. Algorithm~\ref{alg:stream} outlines the general strategy, which can be summarized as follows. First, the baseline solving times of the training instances are determined by solving them with the original, unstreamlined model. These baseline times are later used as reference timeouts.
Then, for about ten minutes, the algorithm repeatedly prompts the LLM to generate candidate streamliners and evaluates their performance on the training instances. Streamliners that cause errors or result in slower solving times than the baseline are discarded, while the promising ones are kept. In the adaptive mode, the evaluation results are also appended as feedback to the prompt, allowing the LLM to refine its suggestions based on past performance. To maintain diversity, every third iteration is performed without feedback, leaving room for new ideas. After the ten minutes have passed, the algorithm selects the best-performing streamliners from all evaluated candidates and outputs them.

Our realtime approach targets scenarios where a user needs to solve
hard constraint problems that typically require hour-long solve
times. Rather than directly attempting these hard instances, the user
provides a few small instances to our system. This proposes efficient
streamliners after interacting with an LLM and a constraint solver
within minutes. The user can then run these streamlined versions of
the original model in parallel with the unstreamlined version,
potentially achieving significant speedup that outweighs the initial
streamliner generation time.


\newcommand{\tbase}{t_\mathrm{base}}

\begin{algorithm}[tb]
\centering
  \caption{General Strategy}
  \begin{algorithmic}[1]
    \REQUIRE an unstreamlined \minizinc model~$\MMM$,
    a set of training instances~$\III$,
    a prompt~$\PPP$,
    an LLM~$\LLL$,
    integers~$n, k$
    \ENSURE a set of streamliners~$\TTT$
    \STATE $\SSS \leftarrow \emptyset$,
           $\SSS' \leftarrow \emptyset$,
           $\PPP' \leftarrow \PPP$
    \STATE Find the baseline solving times~$\tbase(I)$ for each
    instance~$I \in \III$, by solving them with the unstreamlined model~$\MMM$.
    \REPEAT
      \STATE Obtain a set~$\SSS'$ of~$n$ streamliners by prompting the
      LLM~$\LLL$ with prompt~$\PPP'$.
      \STATE Evaluate the performance of the streamliners from~$\SSS'$ in
      parallel on the training instances using the baseline times~$\tbase$ as
      the corresponding timeouts.
      \STATE Discard from~$\SSS'$ the streamliners that produce errors or time
      out on all the training instances.
      \STATE $\SSS \leftarrow \SSS \cup \SSS'$
      \STATE Set $\SSS' \leftarrow \emptyset$ every third
      iteration.\COMMENT{only relevant for \Multishot Mode}
      \IF{\Multishot Mode}
        \STATE Construct the new prompt~$\PPP'$ by appending feedback about the
        streamliners in~$\SSS'$ to the prompt~$\PPP$. This feedback includes
        information such as whether they produced errors or unsatisfiable
        instances; or if their performance was better or worse than the
        unstreamlined model.
      \ENDIF
    \UNTIL 10 minutes have passed.
    \STATE Find a set~$\TTT$ of~$k$ streamliners from~$\SSS$ that collectively
    achieve the best performance over the training instances.
    \RETURN $\TTT$
  \end{algorithmic}
  \label{alg:stream}
\end{algorithm}

\subsection{Prompt Engineering}
\label{sec:prompts}

The prompt given in Figure~\ref{lst:prompt},
is the prompt we use for most of our experiments. In the beginning,
the task is shortly summarized. Then, following the Chain of Thought
technique~\cite{WeiEtal22}, the task is split into single
steps that are explicitly explained. In the end, there are some
compliance rules to ensure the quality and the correct format of the
response. Especially the correct JSON output format is important for
the automated tests on the training instances. We decided on this
prompt because it showed good results in our preliminary experimentation~\cite{voboril_streamllm_2025}.

\newenvironment{promptbox}[1]{
  \centering  
  #1
\vspace{-5pt}    \begin{tcolorbox}[
width=.7\textwidth,
colback=blue!5!white,
colbacktitle=white,
coltitle=black,
left=2pt,
right=2pt,
top=2pt,
bottom=2pt,
boxrule=0.0pt]\parskip=0.2\baselineskip    
}{\end{tcolorbox}}

\begin{listing}[tb!]
\newcommand{\mznconst}{``constraint \texttt{<}\minizinc constraint\texttt{>}''}
\caption{Base Prompt}%
\label{lst:prompt}
\medskip
\begin{promptbox}{Objective}
  Analyze the given MiniZinc code and suggest five additional constraints to enhance the problem-solving process. These constraints can include streamlining, implied, symmetry-breaking, or dominance-breaking constraints.
\end{promptbox}

\begin{promptbox}{Steps}
1. Analyze Content: Read the provided MiniZinc code. Understand the problem being addressed, including its variables, constraints, and optimization goals.

2. Generate additional Constraints: Based on your analysis, create five unique constraints. These should offer targeted modifications or restrictions designed to reduce the search space effectively.

3. Always return your constraints as a JSON object, adhering to the structure: \{``streamliner\_1'': \mznconst, \dots, ``streamliner\_5'': \mznconst\}. Your final output should exclusively be the JSON object containing the five constraints.

4. As a response, you will get feedback for each constraint. Some
constraints might lead to errors, timeouts, or unsatisfiable
instances.

5. Use the information provided in the previous step to generate five new, and hopefully better constraints.

6. Repeat steps 3 to 5 multiple times.
\end{promptbox}

\begin{promptbox}{Compliance Rules}
  1. Response Format: Your final output should exclusively be the JSON object containing the five constraints, adhering to the structure: \{``streamliner\_1'': \mznconst, \dots, ``streamliner\_5'': \mznconst\}. Do not forget the semicolon at the end of the constraint!

  2. Code Quality: All MiniZinc code provided for the constraints must be syntactically correct and functional. For some functions you may need to include an additional library.

  3. Creativity: You're encouraged to be innovative in proposing constraints, keeping in mind their purpose: to narrow down the search space efficiently without oversimplifying the problem.
\end{promptbox}
\end{listing}

\subsection{Obfuscation and Disguise}
\label{sec:obfuscation-and-disguise}

\newcommand{\map}[2]{$\text{#1} \mapsto \text{#2}$}

The capabilities of LLMs to generate original ideas versus relying on
\emph{memorization} remains a key debate in AI
research~\cite{LeeLC023,McCoyEtal23}. This question is particularly relevant
for streamliner generation, as the practical value would be limited if
LLMs could only identify streamliners for well-documented problems
where such techniques are already published and likely included in their
training data. To investigate this capability empirically, we examine two
problem variants: disguised and obfuscated versions.

To \emph{disguise}
a problem, we rename all occurrences of all identifiers such that they
express the semantics of a different problem while maintaining the
syntax and the structure of the original problem. We also insert
comments describing the semantics of the new problem. For instance, to
disguise the well-known \SGfull problem as a newly invented ``Zoo Habitat
Rotation'' problem, we map \map{Golfer}{Animal}, \map{Group}{Habitat},
and \map{Week}{Season}.   Additionally, we add the following
  comment at the top of the MiniZinc model:

\begin{quote}
``The zookeepers at a large zoo are organizing seasonal habitat rotations to
encourage socialization among animals. Each season, animals are grouped into
mixed-species habitats of a fixed size. The zookeepers want to ensure that
no two animals share a habitat more than once during the rotation schedule,
while maximizing the number of season the rotations can last.
This problem seeks for an allocation of animals to habitats which ensures
maximum variety in animal pairings across seasons while maintaining the
integrity of habitat sizes.''
\end{quote}

\emph{Obfuscating} a problem is similar to disguising, except that instead of
mapping to names with the semantics of a different problem, we map to arbitrary
names like id8, id12, and id5, and we strip all comments from the model to
avoid revealing the original problem. Thus, for each original problem in our
benchmark, we introduce two new problems to our benchmark by disguising and
obfuscating the model (\texttt{.mzn} file) and instances (\texttt{.dzn} files)
of the original problem. We then generate and evaluate streamliners for
these problems using the same prompts as the original problem.

\subsection{Combinations of Constraints}
\label{sec:combinations}

Previous work by Spracklen et al.~\shortcite{SpracklenDAM23}
demonstrates that combining
multiple streamlining constraints can be more effective than using them
individually. To evaluate how combinations of streamliners perform in our
\RTSLLM approach, we provide another prompt alternative to the base prompt. It
was created by modifying the LLM base prompt to output combinations of
constraints rather than single constraints, with corresponding adjustments to
the required JSON format. We evaluate the results with this alternative prompt
in Section~\ref{exp:comb}. To establish a baseline for our approach, we
evaluate Spracklen et al.'s streamliner combinations on our benchmark
instances, with detailed results presented in
Subsection~\ref{exp:spracklen}.

\subsection{Offline Approach}

The offline approach investigates whether extended training time
improves streamliner quality. We increase the timeout from 10 minutes to 4
hours and expand the training set from 15 instances solvable within 10 seconds
to 35 instances with solving times between 1 and 30
seconds. The process follows the static approach outlined in
Algorithm~\ref{alg:stream}. In each iteration, we randomly pick GPT-4o or
Claude
3.5 Sonnet, using either the original or combination-generating prompt
described in Section~\ref{sec:combinations}  to leverage the strengths
of different LLMs and prompting strategies.

\section{Experiments}


All instances,  MiniZinc models, and
the Python implementation of \RTSLLM (which includes the prompts) are available on Zenodo\footnote{\url{https://zenodo.org/doi/10.5281/zenodo.13331048}}.

\subsection{Setup and Hardware}

We use MiniZinc~2.8.3 and the Chuffed~0.13.1 solver. As LLMs we use GPT-4o
(\texttt{gpt-4o-2024-11-20}) and Claude~3.5 Sonnet
(\texttt{claude-3-5-sonnet-20241022}) and access them via the openai~1.29.0 and
anthropic~0.25.8 packages in Python~3.11.5. We evaluate the running times of
the test instances on compute nodes with two 2.4~GHz 10-core Intel Xeon
E5-2640~v4 processors with 80~GB of RAM each.

\subsection{Benchmark Problems}

Although our method can be applied to optimization problems, similar to
previous work by Spracklen et al.~\cite{SpracklenDAM23}, we focus here on
decision problems.
This way we can evaluate streamliners only on the basis of their running times.
For optimization problems, one needs to consider solution quality as well.  We
construct a benchmark set of seven constraint satisfaction problems to
test the LLM-generated streamliners. %
\renewcommand*{\thefootnote}{\fnsymbol{footnote}}%
\newcommand{\refcsplib}{\footnotemark[2]}%
\newcommand{\modelcsplib}{\footnotemark[3]}%
\newcommand{\modelmznbench}{\footnotemark[4]}%
\newcommand{\refcarpet}{\footnotemark[1]}%
\footnotetext[2]{Problem description and references available at %
  \url{https://www.csplib.org/Problems}}%
\footnotetext[3]{Model from \url{https://www.csplib.org/Problems}}%
\footnotetext[4]{Model from %
  \url{https://github.com/MiniZinc/minizinc-benchmarks}}%
\footnotetext[1]{Problem described by Schutt et al.~\shortcite{SchuttSV11}}%
This includes four problems which are already considered by
Spracklen et al.~\cite{SpracklenDAM23}, namely
\textbf{\BIBDfull (\BIBD)}\refcsplib\modelcsplib,
\textbf{\CSfull (\CS)}\refcsplib\modelmznbench,
\textbf{\SGfull (\SG)}\refcsplib\modelcsplib,
and
\textbf{\VLfull (\VL)}\refcsplib\modelcsplib.
In addition to that, we include two other well-known problems,
\textbf{\BHfull~(\BH)}\refcsplib\modelmznbench
and
\textbf{\CCfull (\CC)}\refcarpet\modelmznbench.
The main criterion for selecting these six problems is that the \minizinc model
must be readily available. Further filtering is done based on the ease of
generation of instances of the desired difficulty.
Note that, if present, we strip any redundant constraints, symmetry-breaking
constraints, and excessive documentation from the \minizinc models of these
problems. This is the case for \CC and \SG. The only exception is \BIBD, for
which, without symmetry-breaking,
we are unable to find problem instances of the desired difficulty.
\renewcommand*{\thefootnote}{\arabic{footnote}}

Finally, we added the well-known NP-hard \textbf{\HCfull (\HC)}
problem~\cite{Lovasz73,GareyJohnson78}
(described below), for which, to the best of our knowledge, no constraint model
is available. The main reason for including this problem is to see whether our
approach relies on memorization (e.g.,
Lee et al.~\shortcite{LeeLC023} and McCoy et al.~\shortcite{McCoyEtal23}).

\begin{tcolorbox}[size=small]
\myitembegin{\HCfull (\HC):}
  Given integers $c$ and $m$, a set $V$ of vertices, and a set $E$ of hyperedges,
  where each hyperedge is a subset of $V$, find a coloring of vertices using
  at most $c$ colors such that no hyperedge is monochromatic, i.e., is only
  incident to vertices of the same color. Further, the \emph{imbalance}, i.e.,
  the difference in sizes of the largest and the smallest color class, must not
  exceed~$m$.
\end{tcolorbox}




\subsection{Instance Generation}

Since we work with a broad range of problems, working with instances of
similar/uniform difficulty ensures the generality of our findings and prevents
bias from creeping in due to any one particular problem. Hence we construct our
own data set by generating instances of desired difficulties. For each problem,
we create 15 \emph{training instances} that take less than 10 seconds. Further,
we create at least 50 \emph{test instances} that take between 10 minutes and 2
hours to solve.
The 10-minute lower limit is because we want our approach to be comparable in a
realtime setting. The 2-hour upper limit is due to limited computational
resources.


We generated satisfiable instances of the desired difficulties for all the above
problems. Since \SG, \BH, and \BIBD all have relatively simple input
specifications, i.e., a small number of integer parameters, we generated
instances exhaustively and tested their difficulty. For \CC, \CS, \HC, \VL we
generated instances using the AutoIG pipeline developed by Dang et
al.~\shortcite{DangEtal22}. This involved coming up with a parameterized Essence
specification describing the input space of each problem. We then fix a
difficulty window in the form of lower and upper bounds for the \minizinc
running time and then use the automatic algorithm configuration tool
\texttt{irace} to find an optimal configuration of those parameters such that
the constraint-solving tool Conjure can find an instance that is very likely to
fall in the desired difficulty window.

\begin{figure}
  \centering
  \includegraphics[width=0.7\linewidth]{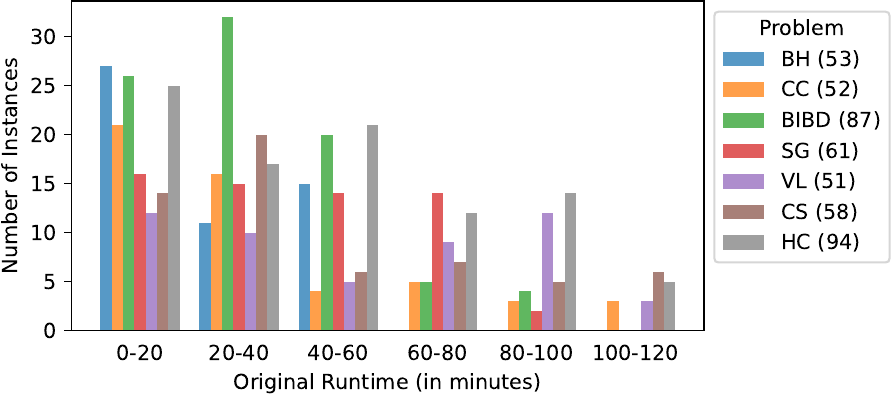}
  \caption{Histogram showing the number of instances for each problem, sorted by
  their original running times and partitioned into 20-minute intervals.
  Further, the total number of instances for each problem is shown next to the
  problem name in the legend.}
  \label{fig:orig-runtime-hist}
\end{figure}

A detailed visualization of the distribution of the original running times (or
in other words, difficulty levels) of the generated instances is shown in
Figure~\ref{fig:orig-runtime-hist}. The total number of instances for each
problem is stated in the legend. Overall, the compiled CSP instances range from
tens of variables and constraints to millions of variables and constraints with
an average of around 200,000 variables and around 300,000 constraints. We list
the ranges of the problem-specific parameters of our benchmark instances below.
\begin{description}
  \item[\BHfull:] Each instance is just a different way of splitting and
    permuting a deck of 52 cards into 3 stacks.
  \item[\CCfull:] Number of room $\in [1, 6]$, number of room rectangles
    $\in [4, 38]$, number of stairs $\in [1, 5]$.
  \item[\BIBDfull:] $v \in [9, 46]$, $k \in [2, 13]$, $\lambda \in [1, 15]$.
  \item[\SGfull:] Number of groups $\in [6, 38]$, number of golfers
    $\in [21, 174]$, number of weeks $\in [2, 27]$.
  \item[\VLfull:] Number of classes $\in [1, 5]$, number of containers
    $\in [19, 67]$, vessel area $\in [153, 512]$.
  \item[\CSfull:] Number of cars $\in [20, 48]$, number of configurations
    $\in [13, 20]$, number of options $\in [3, 6]$.
  \item[\HCfull:] $|V| \in [20, 100]$, $|E| \in [10, 99]$, number of colors
    $\in [3, 5]$.
\end{description}

\subsection{k-Best Selection}
\label{three_best}
As already explained in Section~\ref{sec:streamliners}, there is no
guarantee that every satisfiable instance in the original model is
also satisfiable in a streamlined model. It also might happen that
one streamliner works well on the training instances but is
impractical for the large test instances. In order to get more
robustness, we decided to not only rely on one streamliner but also to
return $k$ streamliners that work best on the training
instances.
In the experiments, we run the original model and $k=3$ streamlined
models in parallel and stop as soon as  any of the $k+1$ models have found a
solution.


\subsection{Preliminary Experiment A}
\label{exp_training_time}

The objective
of this experiment is to determine a suitable maximal running time
$\ttrain$ of the unstreamlined model for the training instances. 
We want to keep $\ttrain$ as low as possible so that the evaluation can be done
quickly but still provide good predictions for the streamliner performance on
the significantly larger test instances.

We run the experiment with the problems \BIBD, \BH, and \SG. For each
problem, we let the LLM suggest ten streamliners. Then, we pick those
three streamliners that perform best on the training instances, and
among the three, we pick the one that performs best on the test
instances.  We use five sets of training instances with
an upper bound of $\ttrain\in \{10, 20, 30, 60\}$ seconds,
respectively. The experiments show that $\ttrain$ has no influence on the
one streamliner that was picked in the end, hence we settled on the
upper bound $\ttrain=10$ for the following experiments.
To reduce the influence of I/O operations, we require training
instances to take at least 1 second to be solved. 


\subsection{Preliminary Experiment B}
\label{exp_training_number}

The goal of this experiment is to decide on the number $\ntrain$ of training
instances. More training instances promise better results on the test instances
but make the evaluation process longer, which we want to avoid for realtime
streamliner generation. Hence we aim at a fair compromise.

\begin{figure}
  \centering
  \includegraphics[width=0.6\linewidth]{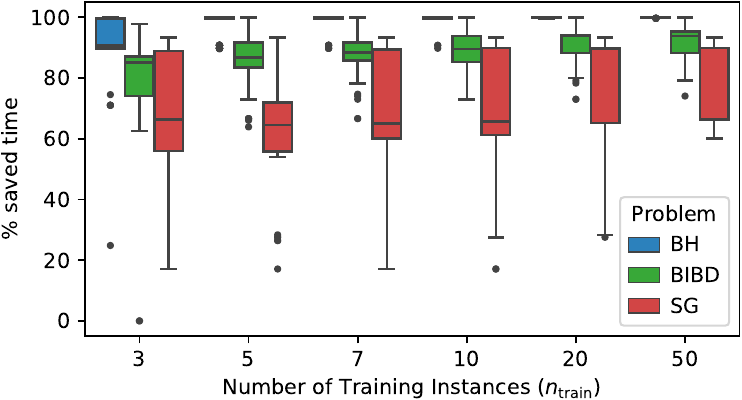}
  \caption{Normalized saved times for different number of training
  instances for three problems}
  \label{fig:combined-boxplot}
\end{figure}

We conduct the experiment with the same three problems as in Experiment~1a. For
each problem, we generate ten streamliners.
For each $\ntrain \in \{3, 5, 7, 10, 20, 50\}$, we randomly pick $\ntrain$
training instances. We determine the combination of three out of the ten
streamliners that perform best on the training instances and compute the time
this combination saves on the test instances, normalized by the time saved by
the virtually best combination of three streamliners. We repeat this  100 times.
The box plots in Figure~\ref{fig:combined-boxplot} shows the normalized saved
running times for different $\ntrain$. As expected, the larger $\ntrain$, the
more time is saved; setting $\ntrain=15$ seems a fair compromise for the
forthcoming experiments.

\subsection{Base Realtime Approach}
\label{sec:base_realtime}

\begin{figure}
  \centering
  \includegraphics[width=\textwidth]{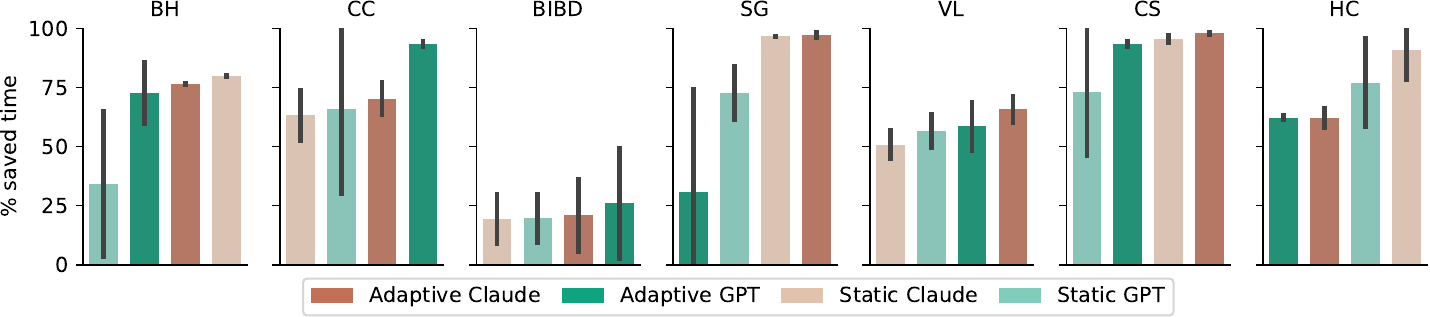}
  \caption{Percentage of reduction in solving time with both \singleshot and
  \multishot approaches using LLMs Claude and GPT across seven problems. Each
  bar and black line denotes, respectively, the mean and standard deviation of
  two runs.}
  \label{fig:saved-time-ortho}
\end{figure}

This is the main experiment where we evaluate the realtime approach.  We run
\RTSLLM  on all seven problems, comparing the \singleshot and
\multishot approach and the two considered LLMs. 
We conducted the experiment twice to fathom the influence of
randomness on the results. Figure~\ref{fig:saved-time-ortho}
reports the average time reduction of both runs, where the time reduction is computed as \texttt{(old runtime - new runtime)/old runtime}. 


Overall, the realtime approach achieves very encouraging results. Some runs
achieve a reduction in running times of 98\% and more. This is the case for the
\CS, the \SG, and the \HC problem. In some cases, the streamlined models could
even solve most of the test instances in less than a second. The good
performance on the \HC problem suggests that the LLMs do not just copy and paste
known streamliners from literature, but are also capable of dealing with new
problems.
For a few of the problems, most noticeably the \BIBD problem, the reduction is
only moderate. This is not so surprising considering that the model for \BIBD
already includes symmetry-breaking constraints and so it turns out to be more
challenging to find significant time savings. It is worth noting that \BIBD also
sees very minor improvement in the approach by Spracklen et
al.~\shortcite{SpracklenDAM23}.


Overall, Claude works slightly better than GPT. Comparing the \multishot and
\singleshot approaches, there is no clear winner. This might be because the
\multishot feedback reduces the exploratory potential of the LLMs.

\begin{figure}
  \centering
  \includegraphics[width=0.75\linewidth]{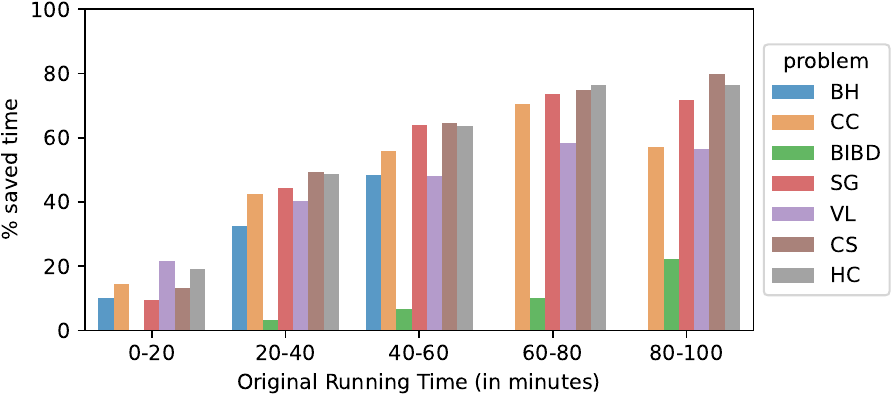}
  \caption{Solving time reduction with respect to the original solving time
    when including streamliner generation time in the total streamlined running
    time.}
  \label{fig:original-vs-saved-time}
\end{figure}

Our analysis thus far compared the time reduction of streamlined models over the
unstreamlined model, but did not consider the time \RTSLLM spent on streamliner
generation. To revisit the realtime scenario as sketched in the introduction,
\Cref{fig:original-vs-saved-time} shows the percentage of saved time when
including the streamliner generation time in the running time of the streamlined
model\footnote{Since there are only a few instances where the unstreamlined
model takes between 100 and 120 minutes, we disregarded this interval.}. As
expected, the saved time for instances that take less than 20 minutes is
relatively poor since more than half of the time is spent on streamliner
generation. For larger instances, however, the generation time is relatively
insignificant, and \RTSLLM shows a remarkably strong performance on most of the
problems, with the exception of the \BIBD problem. 
Note that, the \BH problem has no instances that take more than 60 minutes.

\subsection{Obfuscation and Disguise}

\begin{figure}
  \centering
  \includegraphics[width=\linewidth]{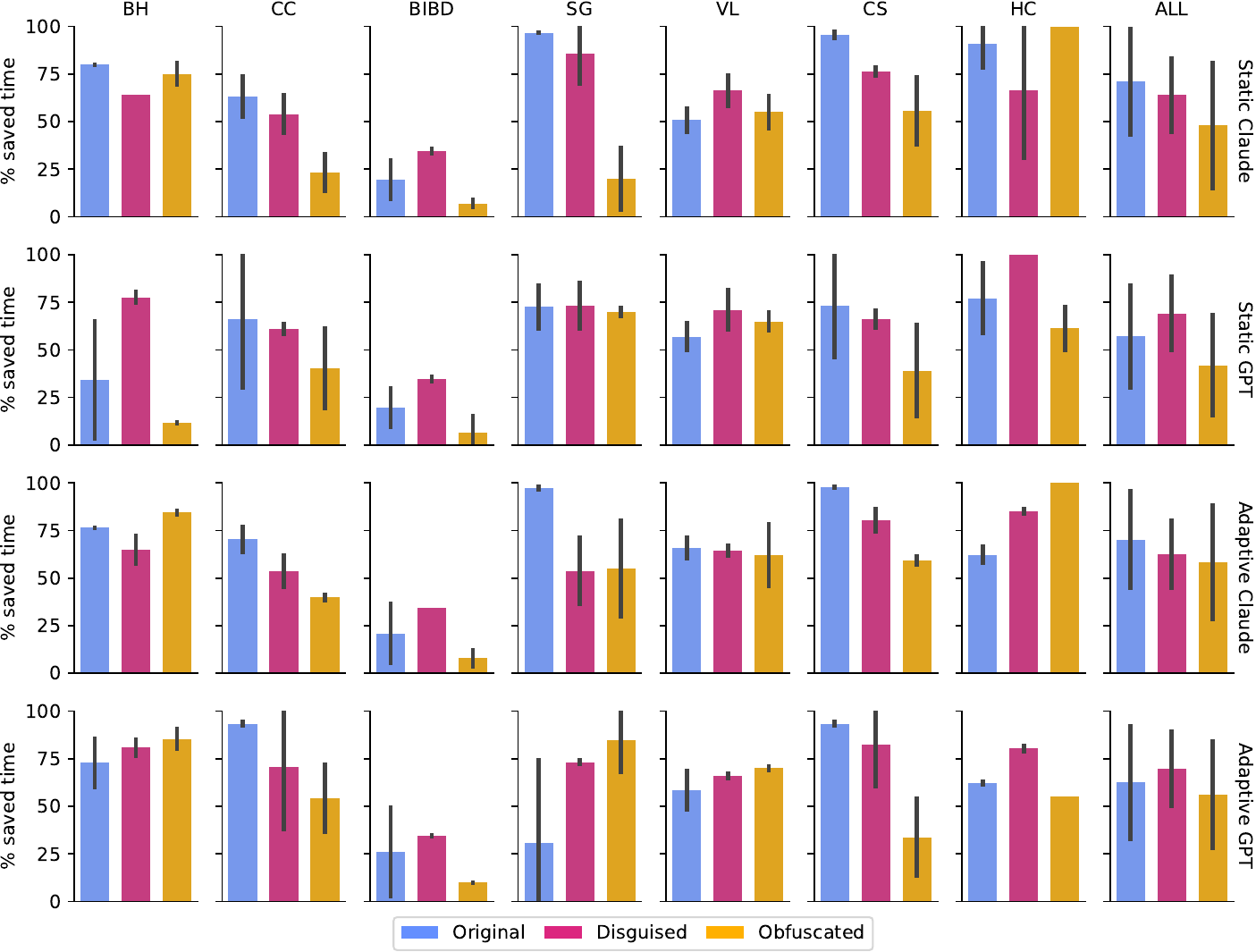}
  \caption{Percentage of reduction in solving time for all seven original
  problems, as well as their disguised and obfuscated versions, with both
  \singleshot and \multishot approaches using LLMs Claude and GPT. The bars and
  black lines denote the means and standard deviations, respectively.}
  \label{fig:obfuscation-disguise}
\end{figure}

This experiment evaluates how obfuscation and disguise affect the performance of
our \RTSLLM approach. Figure~\ref{fig:obfuscation-disguise} shows the results.
It is interesting to see is that for GPT, our approach performs, on average,
slightly better on disguised problems than on the original problems. For Claude,
it performs slightly better on the original problems. In any case, the good
results on the disguised problems suggest that the LLMs do not only copy and
paste online available streamliners for known problems but can make sense of the
underlying problem and can find streamliners for new problems. For the
obfuscated problems, our approach performs, on average, worse than on the
original and disguised problems. This suggests that LLMs prefer semantically
richer language over pure abstract terminology to make better sense of variables
and constraints. 

\subsection{Combinations}
\label{exp:comb}
\begin{figure}
  \centering
  \includegraphics[width=0.7\linewidth]{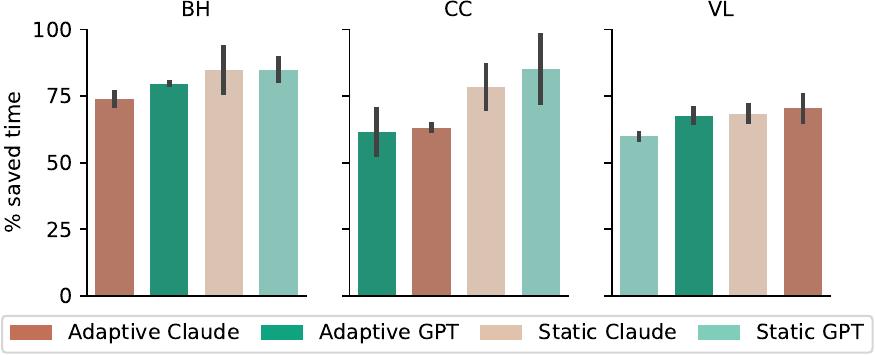}
  \caption{Percentage of reduction in solving time obtained by combinations of
  streamliners with both \singleshot and \multishot approaches using LLMs Claude
  and GPT. The bars and black lines denote the means and standard deviations,
  respectively.}
  \label{fig:combination}
\end{figure}

We evaluate streamliner combinations on the problems \BH, \CC, and
\VL, which we selected because they exhibited moderate performance
improvements with individual streamliners, neither the dramatic gains
seen in problems like \SG nor the minimal improvements observed with
\BIBD. As shown in Figure~\ref{fig:combination}, combining streamliners proves
more effective than individual constraints for \BH
and \VL while slightly reducing performance for~\CC.

\subsection{Comparison with Previous Work}
\label{exp:spracklen}


We replicate a subset of Spracklen et al.'s~\shortcite{SpracklenDAM23}
experiments on our benchmark instances to establish a baseline for comparing our
base approach against. However,
this comparison should be interpreted cautiously due to the following
significant differences between the approaches:
(i)~the technical pipelines differ---their method uses Conjure and Essence for
solving and modeling, whereas we employ MiniZinc for both purposes;
(ii)~the benchmark problems used in their experiments differ from our benchmark
set.


We used the top-performing streamliners from the repository published by
Spracklen et
al.\footnote{\url{https://github.com/stacs-cp/automated-streamliner-portfolios}}
as the initial candidate pool and evaluated them on our benchmark
instances. \BIBD, \CS, \SG, and \VL are the only problems that appear in both
benchmarks. However, the Conjure model for \VL is significantly different
from our \minizinc model, making it incompatible with our input instances.
Thus, we only compare the remaining three problems. We first evaluated the
initial candidates on our training instances to determine the three best
streamliners for each problem, which we then evaluated on the test instances.
The best triples saved 3\%, 17\%, and 21\% time for the \BIBD, \CS,
and \SG problems, respectively. Only three of the nine streamliners from these
triples could satisfy all the test instances.

We see two reasons for these savings being much lower than those reported by
Spracklen et al.: (i) we consider all instances towards the total savings,
while they only average over the improved instances; (ii) the difficulty of our
benchmark instances was calibrated for \minizinc, which need not necessarily
align with the difficulty for Conjure.
Finally, it is also worth noting, for pragmatic reasons, we compared their
approach against our base approach, which operates in real time and does not
generate and train streamliners offline like their method.
Furthermore, their approach allows streamliners to be combinations of
constraints, while our base approach considers only individual constraints.

\subsection{Offline Approach}

\begin{figure}
  \centering
  \includegraphics[width=0.85\linewidth]{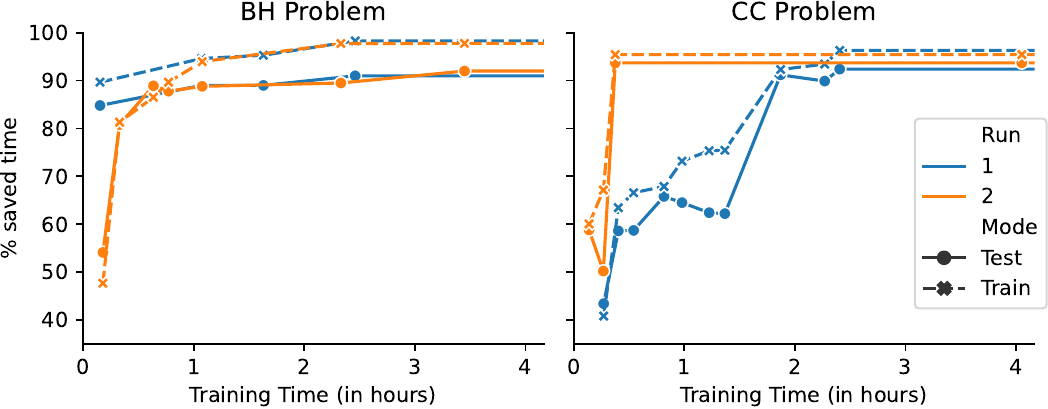}
  \caption{Percentage of reduction in solving time by the offline approach on
  both training and test instances with respect to the length of the training
  phase for two runs each on two problems.}
  \label{fig:offline-approach}
\end{figure}

We conduct the offline approach on the \BH and \CC problems, which show a
middle-ground performance in the main experiment, running each problem twice to
assess the impact of randomness. Figure~\ref{fig:offline-approach} presents the
performance of the current best triple during training and its corresponding
performance on test instances. The strong correlation between training and test
instance trends validates our approach of inferring streamliner quality from
smaller training instances. The performance graph shows significant quality
improvements in the early stages before gradually leveling off. By the end of
training, all runs achieve triples of streamliners that save over 90\% of
running time on test instances, demonstrating that this offline scenario with
extended training times yields better results than the realtime approach. Among
the top triples from each run, combinations make up slightly more than half of
the streamliners, with Claude generating two-thirds and GPT-4 contributing
one-third. This distribution suggests that the diversity achieved through
multiple prompts and LLMs enhances overall results.

\section{Analysis of the Streamliners}

In this section, we analyze the relationship between the individual
streamliners within the best-performing triple for each problem. These triples
are selected based on their performance in the main experiment outlined in
Section~\ref{sec:base_realtime}. Table~\ref{tab:analysis_triples} shows the contribution of each of the
streamliners and the original model, i.e., the percentage of test instances it
performs best. The table also shows the percentage of unsatisfiable test
instances for each streamliner. Further, one can see whether a streamliner is
an implied constraint, a symmetry-breaking constraint, or does not preserve
satisfiability.

It is interesting to see that for the different problems, the contribution of
streamliners is distributed very differently. For the \HC problem, one of the
streamliners contributes to 98\% of the instances, while the other two have
almost no impact. In contrast to this, for the \VL problem, the contribution of
the three streamliners, as well as the original model, are distributed almost
evenly. Further, one can see that for some of the problems, the original model
has no contribution at all, which means that for each instance there is at
least one of the streamliners in the triple that performs better than the
original model.

It is worth noting that for some of the triples, the streamlined models make
none or only a few of the instances unsatisfiable, whilst others have rather
high percentages of unsatisfiable instances. Especially for the \BH and the
\BIBD problem, even the streamliner with the highest contribution makes about
one-third of the instances unsatisfiable.
Of the 21 total streamliners mentioned in Table~\ref{tab:analysis_triples}, 10
streamliners managed to satisfy all the test instances. We further manually
inspected these 10 streamliners to find that 7 of them are indeed solution
preserving, i.e., either implied or symmetry-breaking.
All the 21 streamliners are listed in the appendix.

    
\setlength{\tabcolsep}{13pt} 
\begin{table}
\centering
\caption{Percentages of contribution and percentage of unsatisfiable instances
for all streamliners constituting the best-performing triples for all problems.
The type column indicates whether the streamliner is an implied constraint (i),
a symmetry-breaking constraint (s), or a constraint that does not preserve
satisfiability (-). A (u) denotes cases where the classification remains uncertain.}
\label{tab:analysis_triples}
\begin{tabular}{lrrrrrrrccc}
\toprule
Problem & \multicolumn{4}{c}{Contribution (\%)} &
  \multicolumn{3}{c}{Unsat Instances (\%)} & \multicolumn{3}{c}{Type}
  \\ \cmidrule(lr){2-5} \cmidrule(lr){6-8} \cmidrule(l){9-11}
  & original & s1 & s2 & s3 & s1 & s2 & s3 & s1  & s2  & s3  \\ \midrule
BH   & 11 & 60 & 26 &  2 & 32 & 13 & 77 &  -  &  -  &  -  \\
CC   &  0 & 85 & 12 &  4 &  2 &  2 & 44 &  -  &  -  &  -  \\
BIBD &  7 & 56 & 31 &  6 & 36 &  0 & 79 &  -  &  u  &  -  \\
SG   &  0 & 74 & 25 &  2 &  0 &  5 &  0 &  s &  -  &  s \\
VL   & 25 & 29 & 24 & 22 &  0 &  0 &  0 &  i  &  i  &  i  \\
CS   &  0 & 84 & 12 &  3 &  0 &  5 &  0 &  i  &  -  &  u  \\
HC   &  0 & 98 &  1 &  1 &  0 &  0 &  1 &  i  &  u  &  -  \\
\bottomrule
\end{tabular}
\end{table}

To complement this numerical analysis, we now present and discuss a few of the streamliners covered in the table.

\paragraph{\BHfull (\BH)}
\begin{description}
  \item[s1:] \texttt{constraint x[26] == 26;}
\end{description}

This streamliner for the \BHfull problem fixes the card with ID 26 to appear in the 26th position of the play sequence. It reduces the search space and allows about 60\% of our instances to be solved more efficiently. However, adding this constraint does not preserve satisfiability and makes about one-third of our instances unsatisfiable.

\paragraph{\SGfull (\SG)}
\begin{description}
  \item[s1:] \texttt{constraint forall(g in Golfer)
    (assign[g,2] = ((g-1) + n\_per\_group) mod n\_groups + 1);}
  \item[s3:] \texttt{constraint forall(g in Golfer)
    (assign[g,1] = (g-1) div n\_per\_group + 1);}
\end{description}

The first shown symmetry-breaking constraint, s1, assigns each golfer to a specific group in week two, while s3 assigns each golfer to another specific group in week one. These constraints fix certain elements of the \texttt{assign} array and thereby narrow the search space. Note that s3 is a standard symmetry‑breaking constraint that appears in online resources~\footnote{\url{https://www.csplib.org/Problems/prob010/models/social_golfers1.mzn.html}}. Because of the promising performance of s1, we had a closer look at it and tested whether it is also generalizable to larger instances. The results are remarkable: We came across many instances that could not be solved by the original model within 10 hours, but were solved by the streamlined model within 3 minutes. This corresponds to a runtime reduction of more than 99\%.

\paragraph{\HCfull (\HC)}
\begin{description}
  \item[s1:] \texttt{constraint sum(c in Color)
    (num\_vertices\_of\_color[c]) = num\_vertices;}
\end{description}

This constraint for the \HCfull problem ensures that the sum of the vertices of every color equals the total number of vertices in the graph. Although it is already implied by other constraints in the encoding, its explicit inclusion significantly improves the solver performance.

\section{Conclusion and Future Work}
Our \RTSLLM system and its analysis demonstrate that LLMs can
effectively generate streamlining constraints for constraint
satisfaction problems. Our experiments across seven diverse problems
show impressive runtime reductions compared to unstreamlined models.
The system can generate effective streamliners within minutes using
only small training instances, making it practical even for problems
with hour-long solving times. \RTSLLM performs significantly better
on the original and the disguised problem variants than on the
obfuscated ones.  This suggests a similarity to human experts, who
naturally leverage the semantic context.

Our work gives rise to several promising research directions and possible extensions of our approach:
(i)~Knowledge distillation could compress the LLM components into smaller,
locally deployable language models specifically tuned for streamliner
generation.
(ii)~The observation that disguised problems sometimes yield better results
suggests systematically exploring problem translations as a meta-optimization
strategy.
(iii)~Developing LLM-supported formal verification methods for streamliner soundness could enable meaningful evaluation on unsatisfiable instances. This would allow us to infer that the original instance itself was unsatisfiable if the streamlined instance yields an unsatisfiable outcome.
And finally, (iv)~combining LLM-generated streamliners with evolutionary
optimization techniques could improve their effectiveness while
maintaining the rapid generation capabilities demonstrated in this
work.

While this study focuses exclusively on constraint satisfaction problems, the general approach can also be applied to optimization problems. In our recent work~\cite{Voboril_CP25}, we applied the approach to optimization problems and achieved results surpassing the previously best-known solutions.

\appendix

\section*{Appendix: Streamliners}

\paragraph{\BHfull (\BH)}
\begin{description}
  \item[s1:] \texttt{constraint x[26] == 26;}
  \item[s2:] \texttt{constraint forall(i in 1..51) (x[i] != x[i+1] + 1);}
  \item[s3:] \texttt{constraint x[52] == 52;}
\end{description}

\paragraph{\CCfull (\CC)}
\begin{description}
  \item[s1:] \texttt{constraint forall(i in Stairs, j in st\_rec\_ids[i])
    (st\_rec\_x[j] mod st\_rec\_len[j] = 0);}
  \item[s2:] \texttt{constraint forall(i in Stairs, j in st\_rec\_ids[i])
    (st\_rec\_y[j] mod st\_rec\_wid[j] = 0);}
  \item[s3:] \texttt{constraint forall(i in Rooms, j in rm\_rec\_ids[i])
    (rm\_rec\_x[j] mod MinRmRecSize = 0);}
\end{description}

\paragraph{\BIBDfull (\BIBD)}
\begin{description}
  \item[s1:] \texttt{constraint forall(i in rows)\\
    (sum(j in cols where j mod 2 = 1) (bool2int(m[i, j])) >= r div 2);}
  \item[s2:] \texttt{constraint forall(i in rows)
    (sum(j in cols) (bool2int(m[i, j])) >= k div 2);}
  \item[s3:] \texttt{constraint forall(i in rows, j in cols where i = j)
    (m[i, j] = false);}
\end{description}

\paragraph{\SGfull (\SG)}
\begin{description}
  \item[s1:] \texttt{constraint forall(g in Golfer)
    (assign[g,2] = ((g-1) + n\_per\_group) mod n\_groups + 1);}
  \item[s2:] \texttt{constraint forall(w in Week, gr in Group)
    (assign[gr, w] = gr);}
  \item[s3:] \texttt{constraint forall(g in Golfer)
    (assign[g,1] = (g-1) div n\_per\_group + 1);}
\end{description}

\paragraph{\VLfull (\VL)}
\begin{description}
  \item[s1:] \texttt{constraint forall(c in Containers)
    (Left[c] <= deck\_width - width[c]);}
  \item[s2:] \texttt{constraint max([Right[c] | c in Containers])
    <= deck\_width;}
  \item[s3:] \texttt{constraint forall(c1, c2 in Containers where c1 < c2)\\
    (Left[c1] <= deck\_width - width[c1]
    \textbackslash/ Left[c2] <= deck\_width - width[c2]);}
\end{description}

\paragraph{\CSfull (\CS)}
\begin{description}
  \item[s1:] \texttt{constraint forall(i in options)\\
    (count(j in 1..n\_cars) (b\_seq\_confs[i,j] = 1)\\
    = sum(c in configurations) (confs[c,i] * n\_cars\_by\_confs[c]));}
  \item[s2:] \texttt{constraint forall(i in 1..n\_cars-1)\\
    (seq\_confs[i] >= seq\_confs[i+1]\\
    -> forall(j in options) (b\_seq\_confs[j,i] >= b\_seq\_confs[j,i+1]));}
  \item[s3:] \texttt{constraint forall(i in 1..n\_cars-2)\\
    (seq\_confs[i] = seq\_confs[i+1] -> seq\_confs[i+1] != seq\_confs[i+2]);}
\end{description}

\paragraph{\HCfull (\HC)}
\begin{description}
  \item[s1:] \texttt{constraint sum(c in Color)
    (num\_vertices\_of\_color[c]) = num\_vertices;}
  \item[s2:] \texttt{constraint forall(c in Color)\\
    (num\_vertices\_of\_color[c]
    <= (num\_vertices div num\_colors) + (max\_imbalance div 2));}
  \item[s3:] \texttt{constraint forall(v1, v2 in Vertex where v1 < v2)\\
    (if exists(e in Hyperedge)
    (incidence[e,v1] = 1 /\textbackslash\ incidence[e,v2] = 1) then\\
    coloring[v1] <= coloring[v2]\\
    endif);}
\end{description}

\section*{Acknowledgments}\enlargethispage*{5mm}
\sloppypar The authors thank Carlos Ans\'{o}tegui for helpful discussions.  
This research is supported by the Austrian Science Funds (FWF),
projects 10.55776/COE12, 10.55776/P36688, and 10.55776/P36420.
Part of this work was carried out while the third author was a
visiting researcher at the Simons Institute for the Theory of
Computing, Berkeley.

\bibliographystyle{named_doi}
\bibliography{literature}

\end{document}